\documentclass[a4paper,12pt]{iopart}
\usepackage[]{amssymb}
\usepackage[]{amsthm}
\usepackage[]{mathrsfs}
\usepackage[]{tensor}
\usepackage[]{bm}
\usepackage[]{eucal}
\usepackage[symbol]{footmisc}
\newcommand{\dd}{\partial}
\newcommand{\nab}[1]{\nabla_{\!#1}}
\newcommand{\qqd}{\ , \quad}
\newcommand{\bc}{\begin{center}}
\newcommand{\ec}{\end{center}}
\newcommand{\be}{\begin{equation}}
\newcommand{\ee}{\end{equation}}

\newcommand{\rr}{\mathbb{R}}

\newcommand{\w}{\wedge}
\newcommand{\df}{\mathrm{d}}
\newcommand{\Lie}{\textrm{\pounds}}
\newcommand{\heq}[1]{\buildrel #1 \over =}

\newcommand{\je}{J_{\mathrm{e}}}
\newcommand{\jm}{J_{\mathrm{m}}}

\theoremstyle{plain} \newtheorem{tm}{Theorem}[]
\theoremstyle{plain} \newtheorem{lm}[tm]{Lemma}
\theoremstyle{definition} \newtheorem{defn}[tm]{Definition}
\newcommand{\btm}{\begin{tm}}
\newcommand{\etm}{\end{tm}}
\newcommand{\blm}{\begin{lm}}
\newcommand{\elm}{\end{lm}}
\newcommand{\bdefn}{\begin{defn}}
\newcommand{\edefn}{\end{defn}}
\newcommand{\prf}[1]{\noindent \textbf{Proof.} #1 \qed}

\begin{document}

\begin{flushright}
ZTF-EP-14-10
\end{flushright}

\title[On the various aspects of electromagnetic potentials]{On the various aspects of electromagnetic potentials in spacetimes with symmetries}

\bigskip

\author{Ivica Smoli\'c}

\address{Department of Physics, Faculty of Science, University of Zagreb, p.p.~331, HR-10002 Zagreb, Croatia}

\ead{ismolic@phy.hr}

\vspace{20pt}

\begin{abstract}
We revise and generalize the properties of the electric and the magnetic scalar potentials in spacetimes admitting a Killing vector field: Their constancy on the Killing horizons, uniqueness of solution for the electromagnetic test fields and the relation between the Bianchi identity and Maxwell's equations. In each of these examples, collinearity of currents with the Killing vector field is shown to be the crucial property.
\end{abstract}


\bigskip


\vspace{20pt}

\section{Introduction}

The canon of no-hair theorems have taught us that black holes themselves can support only simple electromagnetic fields. Nevertheless, realistic astrophysical objects are always ``dirty'' and can be surrounded by \emph{external} fields, such as those produced by plasma accreting onto the black hole. There is a vast literature aimed at the understanding of basic properties of the electromagnetic fields surrounding the black holes, focused on electrostatic problems \cite{CW71,HR73,Linet76,Linet77,Molnar01}, magnetostatic problems \cite{Wald74,Petterson74,Moss11}, formalism of the membrane paradigm \cite{TMP}, force-free magnetospheres \cite{GJ14}, etc. Still, the literature is full of technical gaps, hiding among the results which are unclear about the underlying assumptions, the scope and the generality of the claims.

\bigskip

One important simplification in various models is the assumption about the presence of symmetries. Mere isometry of the underlying spacetime doesn't necessarily imply that the fields in it will respect the same symmetries. This ``symmetry inheritance'' has been analysed for various fields \cite{MW75,Hoen78} and it has been found \cite{Tod06} that it is, for example, true for the non-null electromagnetic fields in the presence of bifurcate Killing horizons. Here we shall focus our discussion on the spacetimes and electromagnetic fields which share the common geometrical symmetries. The aim of this paper is to revise some basic properties of the electromagnetic field in spacetimes admitting (at least one) Killing vector field.

\bigskip

In section 2 we review the basic facts about Maxwell's equations and electro\-mag\-ne\-tic scalar potentials in general relativistic context. In section 3 we generalize proofs for the constancy of the electric and the magnetic scalar potentials over Killing horizons and discuss the underlying assumptions. In section 4 we generalize the uniqueness theorem for the scalar potentials of the electromagnetic test fields. In section 5 we review relations between the Bianchi identity and the equations of motion for matter fields. In section 6 we generalize the derivation of Maxwell's equations from the Bianchi identity to the nonvacuum case. In the final section we make some concluding remarks. It is important to emphasize that, except in the section 4, none of the results presented in the paper depend on particular form of the gravitational field equations and are thus valid beyond general relativity.

\bigskip 

Throughout the paper we assume metric signature convention $(-,+,+,+)$ and natural system of units. All spacetimes are 4-dimensional, connected smooth Lorentzian manifolds. We shall employ abstract index notation \cite{Wald} or ``indexless'' notation \cite{Heusler}, where appropriate. Furthermore, we use equalities of the form
\be
A \heq{S} B
\ee
to indicate that equality of ``$A$'' and ``$B$'' holds (at least) on points of the set $S$. If not otherwise stated, it is assumed that Einstein's equation always contains the cosmological constant term,
\be\label{eq:EE}
R_{ab} - \frac{1}{2}\,R g_{ab} + \Lambda g_{ab} = 8\pi T_{ab} \ .
\ee

\vspace{20pt}

\section{Enter scalar potentials}

The electromagnetic field is described by 2-form $F_{ab}$, which is a solution to Maxwell's equations. To keep our discussion as general as possible we shall assume that apart from the electric current $\je^a$, there is also the magnetic, or ``monopole'' current $\jm^a$ \cite{Shnir}. Maxwell's equations in the presence of both currents can be written compactly, in the language of differential forms, as
\be\label{eq:MaxF}
\df F = 4\pi\,{*\jm} \qqd \df{*F} = 4\pi\,{*\je}
\ee
or, with the abstract index notation, as
\be
\nab{a}F_{bc} + \nab{b}F_{ca} + \nab{c}F_{ab} = 4\pi \jm^d\,\epsilon_{dabc} \qqd \nab{a}F^{ab} = -4\pi \je^b \ .
\ee
Using any auxiliary smooth vector field $X^a$, we can formally define the electric field 1-form $E_a$ and the magnetic field 1-form $B_a$ (sometimes referred to as the electric and the magnetic ``components'' of $F_{ab}$) with respect to $X^a$ via 
\be
E(X) \equiv -i_X F \qqd B(X) \equiv i_X {*F} \ .
\ee
It is not difficult to prove that we can make the following decomposition
\be\label{eq:Fdecom}
-NF = X \w E(X) + *(X \w B(X))
\ee
where $N = (X|X)$. For example, if $X^a = u^a$ is the 4-velocity of the observer, then $E(u)$ and $B(u)$ are simply conventional electric and magnetic fields measured in the observer's rest frame \cite{KS}. Let us now assume that spacetime $(M,g_{ab})$ admits a Killing vector field $\xi^a$ with the norm $N = (\xi|\xi)$ and the twist $\omega_a$, defined as
\be
\omega = -*(\xi \w \df\xi) \quad \textrm{or} \quad \omega_a = \epsilon_{abcd}\,\xi^b\nabla^c\xi^d \ .
\ee
Along the following discussions, we shall assume that $N \ne 0$ and then separately make comments about the points where $N = 0$. Except in the section 4, none of the results presented here depend on the \emph{sign} of the function $N$. This means that these theorems can be applied in regions of the spacetime where the vector field $\xi^a$ has different causal character, such as the ergoregions or the interiors of the black holes. In this context, it is convenient to define the electric field $E_a \equiv E_a(\xi)$ and the magnetic field $B_a \equiv B_a(\xi)$ with respect to the vector field $\xi^a$. Let us now recast Maxwell's equations (\ref{eq:MaxF}) into equations for the electric and the magnetic field.

\bigskip

\blm
Maxwell's equation (\ref{eq:MaxF}) can be rewritten as a system of differential equations for electric and magnetic 1-forms,
\be\label{eq:MaxEB1}
\delta E - \frac{1}{N}\,\Big[ (\df N|E) + (\omega|B) \Big] = -4\pi (\xi|\je)
\ee
\be\label{eq:MaxEB2}
\delta B - \frac{1}{N}\,\Big[ (\df N|B) - (\omega|E) \Big] = -4\pi (\xi|\jm)
\ee
\be\label{eq:MaxEB3}
\df E = 4\pi *(\jm \w \xi)
\ee
\be\label{eq:MaxEB4}
\df B = -4\pi *(\je \w \xi)
\ee
where $\delta$ denotes the coderivative, $\delta = - {*\,\df\,*}$.
\elm

\medskip

\prf{
Using basic identities from the Appendix A, we have
\be
\delta E = {*\,\df\,*}\,i_\xi F = -*\df({*F} \w \xi) = -i_\xi\,{*\,\df\,*}F - *({*F} \w \df\xi)
\ee
and
\be
\delta B = -{*\,\df\,*}\,i_\xi {*F} = -*\df(F \w \xi) = -i_\xi\,{*\,\df} F - *(F \w \df\xi) \ .
\ee
Furthermore, making use of the decomposition (\ref{eq:Fdecom}) we have
\be
-N *({*F} \w \df\xi) = -(\xi \w E | \df\xi) - *(B \w *\omega) = (E|\df N) + (B|\omega)
\ee
and
\be
- N *(F \w \df\xi) = *(E \w *\omega) - (\df\xi|\xi \w B) = -(\omega|E) + (\df N|B) \ .
\ee
Putting all this together it is straightforward to derive first two Maxwell's equations in the form (\ref{eq:MaxEB1})--(\ref{eq:MaxEB2}) from the original ones (\ref{eq:MaxF}). Finally, using the fact that Lie derivative with respect to a Killing vector commutes with Hodge dual and the identity (\ref{eq:iXsalpha}), we have
\be
\df E = -\df i_\xi F = -\Lie_\xi F + i_\xi\,\df F = 4\pi\,i_\xi {*\jm} = 4\pi *(\jm \w \xi)
\ee
and
\be
\df B = \df i_\xi {*F} = \Lie_\xi {*F} - i_\xi\,{\df\,*} F = - 4\pi\,i_\xi {*\je} = -4\pi *(\je \w \xi) \ .
\ee
}

\bigskip

General analysis and the solution of Maxwell's equations are simplified if it is possible to replace electric and magnetic 1-forms with corresponding scalar potentials. However, these cannot always be defined, as explicated in the following result, a corollary to Lemma 1.

\bigskip

\blm
Let $F_{ab}$ be an electromagnetic field, invariant under the action of a Killing vector field $\xi^a$. Then 

\begin{itemize}
\item[\textnormal{e)}] the electric field $E = -i_\xi F$ is a closed form if and only if the magnetic current $\jm$ satisfies the condition
\be\label{eq:ccm}
\jm \w \xi = 0 \ ;
\ee
\item[\textnormal{m)}] the magnetic field $B = i_\xi {*F}$ is a closed form if and only if the electric current $\je$ satisfies the condition
\be\label{eq:cce}
\je \w \xi = 0 \ .
\ee
\end{itemize}
\elm

\bigskip

If the conditions from the previous lemma are met, then the Poincar\'e lemma allows us to define, at least locally, electric scalar potential $\Phi$ and magnetic scalar potential $\Psi$,
\be
E = \df\Phi \qqd B = \df\Psi \ .
\ee
Intuitively, conditions (\ref{eq:ccm})--(\ref{eq:cce}) imply that the currents $\je^a$ and $\jm^a$ are ``parallel'' to the Killing vector field $\xi^a$. For example, at all points where $N \ne 0$ contraction with $\xi^a$ implies
\be\label{eq:jxijxi}
\je^a = \frac{(\xi|\je)}{N}\,\xi^a \quad \textrm{and} \quad \jm^a = \frac{(\xi|\jm)}{N}\,\xi^a \ .
\ee
Assuming that the current conditions are satisfied, Maxwell's equations for scalar potentials are given by
\begin{eqnarray}
\Delta \Phi - \frac{1}{N}\,\Big[ (\df N|\df\Phi) + (\omega|\df\Psi) \Big] & = -4\pi (\xi|\je) \label{eq:Maxspot1}\\
\Delta \Psi - \frac{1}{N}\,\Big[ (\df N|\df\Psi) - (\omega|\df\Phi) \Big] & = -4\pi (\xi|\jm) \label{eq:Maxspot2}
\end{eqnarray}
Usually we refer to solutions with $\Psi \equiv 0$ as ``purely electric'' and solutions with $\Phi \equiv 0$ as ``purely magnetic''.

\bigskip

Just as in the case of classical electrostatics and magnetostatics \cite{Griffiths}, introduction of the scalar potentials is an important tool in the analysis of the black hole electrodynamics. In fact, these scalars can be interpreted in a much broader sense. If we introduce a 2-form $F = \df\xi$ for a Killing vector field $\xi^a$, then it satisfies Maxwell's equations
\be
\df F = 0 \qqd \df\,{*F} = 4\pi\,{*J_\xi} \ ,
\ee
with ``electric current''
\be
J_\xi \equiv \frac{1}{2\pi}\,R(\xi) \ ,
\ee
where $R(\xi)$ is the 1-form $R(\xi)_a \equiv R_{ab}\xi^b$. Current condition (\ref{eq:cce}) in this case is equivalent to the condition $R(\xi) \w \xi = 0$, and corresponding scalar potentials are the real and the imaginary part of the Ernst potential \cite{Heusler}.

\vspace{20pt}

\section{Killing horizons as equipotential hypersurfaces}

We immediately see that the scalar potentials are constant along the orbits of the Killing vector field $\xi^a$,
\be
\Lie_\xi \Phi = i_\xi \df\Phi = i_\xi E = 0 \qqd \Lie_\xi \Psi = i_\xi \df\Psi = i_\xi B = 0 \ .
\ee
It is a remarkable fact that black hole horizons are, just like conducting surfaces, equipotentials for the scalar potentials. This piece of information is very important in the derivation of the generalized Smarr formula \cite{C73}, the first law of black hole thermodynamics \cite{C73,GaoWald01}, as well as in the black hole uniqueness theorems \cite{Heusler}. Depending on the motivation of the research, this ``zeroth law'' of the black hole electrodynamics can be proven for a class of solutions of the chosen equations of motion (independently of the symmetries), or for a class of spacetimes sharing the same symmetries (independently of the equations of motion). A short review of these approaches was already given in \cite{ISm1}, and here we shall briefly repeat the main results without the proofs.

\bigskip

Carter \cite{C73} has exploited the fact that the contraction $R(\xi,\xi) = R_{ab}\xi^a\xi^b$ vanishes on black hole horizons, and then established constancy via Einstein's equation.

\bigskip

\btm
Let $(M, g_{ab}, F_{ab})$ be a solution of the Einstein--Maxwell equations with a Killing
horizon $H[\xi]$, generated by the Killing vector field $\xi^a$, and electromagnetic field $F_{ab}$, invariant
under the action of the Killing vector field $\xi^a$ and nonsingular on $H[\xi]$. Then the electric and
the magnetic scalar potential, $\Phi$ and $\Psi$, are constant over each connected component of the Killing horizon $H[\xi]$.
\etm

\bigskip

A Killing horizon $H[\xi]$, a basic ``model'' for the black hole event horizons in a presence of spacetime isometries, can be roughly defined as a null hypersurface to which the Killing field $\xi^a$ is normal (for more precise definitions see e.g.~\cite{Heusler,HCC}). The bifurcation surface is a locus of points on the Killing horizon $H[\xi]$ where the generating Killing vector field $\xi^a$ vanishes. Its presence can considerably simplify proofs of many black hole properties, such as the zeroth law of black hole thermodynamics \cite{KayWald91}. This approach has been used by Gao in \cite{Gao03}.

\bigskip 

\btm
Let $(M, g_{ab}, F_{ab})$ be a spacetime which admits a Killing vector field $\xi^a$ and contains
a bifurcate Killing horizon $H[\xi]$. Furthermore, let $F_{ab}$ be a electromagnetic field, invariant under the action of 
the Killing vector field $\xi^a$ and nonsingular on $H[\xi]$. Then the electric and the magnetic
scalar potential, $\Phi$ and $\Psi$, are constant over each connected component of the Killing horizon $H[\xi]$.
\etm

\bigskip

However, assumption of the existence of the bifuraction surface has its limitations since, for example, we already know that the extremal black holes are not of the bifurcate type. In an attempt to fill this technical gap and find the proof which makes no use of gravitational field equations, nor the assumption about the existence of a bifurcation surface, another one has been proposed in \cite{ISm1}. We shall now revise and generalize this result through two new theorems.

\bigskip

Static spacetimes admit a stationary Killing vector $k^a$, which is hypersurface orthogonal, that is $\omega_a = 0$. An immediate consequence is that Maxwell's equations (\ref{eq:Maxspot1})--(\ref{eq:Maxspot2}) become decoupled system of differential equations.
Apart from stationarity, $\Lie_k F = 0$, there are two usual staticity conditions imposed on the electro\-mag\-ne\-tic field in this context,
\be\label{eq:statcond}
F \w k = 0 \quad \textrm{and} \quad {*F} \w k = 0 \ ,
\ee
where the first one corresponds to the purely electric and the second to the purely magnetic case. In fact, earlier equations (\ref{eq:MaxEB3}) and (\ref{eq:MaxEB4}) can be written as
\be\label{eq:deltaFk}
\delta (F \w k) = -4\pi \je \w k \qqd \delta ({*F} \w k) = 4\pi \jm \w k \ ,
\ee
so as to to reveal the close relation between the staticity and the current conditions. Trivially, (\ref{eq:statcond}) imply (\ref{eq:ccm})--(\ref{eq:cce}) via (\ref{eq:deltaFk}), but the converse is not true in general and the sufficient conditions were found by Carter in \cite{C73,C87}. However, his analysis make use of Einstein's equations, which we want to avoid, and the constancy of scalar potentials on the black hole horizon, which we want to prove in the first place. For this reason we shall economise the choice of assumptions as follows.

\bigskip

\btm
Let $(M,g_{ab},F_{ab})$ be a static spacetime with Killing vector field $k^a$, containing a Killing horizon $H[k]$ and electromagnetic field $F_{ab}$ which is invariant under the action of $k^a$, nonsingular on $H[k]$ and which satisfies one of the staticity conditions (\ref{eq:statcond}). Then the electric scalar potential $\Phi$ and magnetic scalar potential $\Psi$ are constant on each connected component of the Killing horizon $H[k]$. 
\etm

\medskip

\prf{
The assumptions from the theorem assure that current conditions are satisfied and therefore scalar potentials are at least locally well defined. If the first of the staticity conditions (\ref{eq:statcond}) is satisfied, $\Psi$ is constant everywhere, and the contraction with $k^a$ gives the relation
\be
0 = i_k (F \w k) = -E \w k + NF \heq{H} -\df\Phi \w k
\ee
From here it follows that
\be
i_Z \df\Phi \heq{H} 0
\ee
for all tangent vectors $Z^a \in TH[k]$ (note that by definition we have $\df\Phi = 0$ at points where $k^a = 0$). Therefore, $\Phi$ is constant over the horizon $H[k]$. In case when the second of the staticity conditions (\ref{eq:statcond}) is satisfied, the proof for the constancy of $\Psi$ is completely analogous.  
}

\bigskip

Now we turn our attention to the stationary axisymmetric spacetimes $(M,g_{ab})$ with the corresponding commuting Killing vectors, stationary $k^a$ and axial $m^a$ (with closed orbits). We shall adopt conventional notation for the inner products between these Killing vectors,
\be
V \equiv -(k|k) \qqd X \equiv (m|m) \qqd W \equiv (k|m) \ .
\ee
We assume the absence of the closed causal curves in the exterior of the horizon, so that $X \ge 0$, with the equality holding only on the rotation axis (where $m^a$ vanishes). If, in addition, hypersurfaces orthogonal to these Killing vector fields are integrable,
\be
k \w m \w \df k = k \w m \w \df m = 0 
\ee
then we say that the spacetime is \emph{circular}. Suppose now that for the vector field
\be
\xi^a = k^a + \Omega\,m^a \qqd \Omega = -W/X \ ,
\ee
we define the hypersurface
\be
S[\xi] = \left\{ p \in M \, : \, (\xi|\xi)|_p = 0 \right\}
\ee
in the circular spacetime $(M,g_{ab})$. Then the weak rigidity theorem (\cite{Heusler}, Theorem 8.13) claims that $\Omega$, the ``angular velocity of the horizon'', is constant on $S[\xi]$, $\xi^a$ is a Killing vector field at least on $S[\xi]$ and $S[\xi]$ is a null hypersurface, hence a Killing horizon. In order to emphasize this particular Killing vector field, we shall use special notation,
\be
\chi^a \equiv k^a + \Omega_H m^a \qqd \Omega_H \equiv -\frac{W}{X}\Big|_H \ .
\ee 
Suppose that circular spacetime contains a Killing horizon $H[\chi]$, invariant under the action of the Killing vector fields $k^a$ and $m^a$. This means that $k^a$ and $m^a$ are tangent to $H[\chi]$ and hence
\be
(\chi|k) \heq{H} 0 \quad \textrm{and} \quad (\chi|m) \heq{H} 0 \ ,
\ee
from where it follows that
\be
\Omega_H = \frac{V}{W}\Big|_H \ .
\ee 
If $j^a$ is either electric or magnetic current, then the current condition $j \w \chi = 0$ (here we assume that $\Omega_H \ne 0$) implies ``circularity'' of the current, $j \w k \w m = 0$. In what follows, we shall need the generalized version of the Proposition 5.6 from \cite{Heusler}.

\bigskip

\blm
Let $(M,g_{ab},F_{ab})$ be a stationary axisymmetric spacetime with commuting Killing vectors $k^a$ and $m^a$, containing a stationary axisymmetric electromagnetic field $F_{ab}$, $\Lie_k F = \Lie_m F = 0$, and electric and magnetic currents which satisfy circularity conditions,
\be
\je \w k \w m = \jm \w k \w m = 0
\ee
Then, in every domain of spacetime which intersects the rotation axis, Maxwell's equations imply 
\be
F(k,m) = -*(k \w m \w {*F}) = 0 \qqd {*F}(k,m) = *(k \w m \w F) = 0 \ .
\ee
\elm

\bigskip

\prf{
Using Cartan's identity (\ref{eq:Cartan}) and commuting of the two Killing vectors, $\Lie_k m = 0$, we have identity
\be
\df i_m i_k = i_k \Lie_m - i_m \Lie_k + i_m i_k \df
\ee
which can be applied on $F_{ab}$ and ${*F}_{ab}$, with help of Maxwell's equation,
\be
\df F(k,m) = (i_k \Lie_m - i_m \Lie_k) F + 4\pi *(\jm \w k \w m)
\ee
\be
{\df\,*}F(k,m) = (i_k \Lie_m - i_m \Lie_k)\,{*F} + 4\pi *(\je \w k \w m) \ .
\ee
Using these equations and the assumptions about symmetry of the field, it follows that scalars $F(k,m)$ and $*F(k,m)$ are constants. Moreover, since $F(k,m) = 0$ and ${*F}(k,m) = 0$ on the rotation axis (where $m^a = 0$), it follows that $F(k,m) = 0 = *F(k,m)$ in every domain of spacetime which intersects the rotation axis. 
}

\vspace{25pt}

\noindent
It is important to note that in order to establish the validity of restricted equations,
\be
F(k,m) \heq{H} 0 \qqd {*F}(k,m) \heq{H} 0 \ ,
\ee 
it is enough to assume circularity of the currents on the horizon $H$, 
\be
\je \w k \w m \heq{H} 0 \qqd \jm \w k \w m \heq{H} 0 \qqd
\ee
provided that the rotation axis intersects the horizon $H$.

\bigskip

\btm
Let $(M,g_{ab},F_{ab})$ be a stationary axisymmetric spacetime containing a Killing horizon $H[\chi]$ and a stationary axisymmetric eletromagnetic field $F_{ab}$, nonsingular on $H[\chi]$, which is a solution to Maxwell's equations (\ref{eq:MaxF}) with currents which satisfy circularity conditions
\be
\je \w \chi \heq{O} 0 \quad \textrm{and} \quad \jm \w \chi \heq{O} 0
\ee
on some open set $O \supset H[\chi]$. Then the electric scalar potential $\Phi$ and the magnetic scalar potential $\Psi$ are constant on each connected component of the Killing horizon $H[\chi]$. 
\etm

\medskip

\prf{
The assumptions from the theorem assure that scalar potentials are at least locally well defined and, by the result of previous lemma, we have
\be
k \w m \w F \heq{H} 0 \quad \textrm{and} \quad k \w m \w *F \heq{H} 0 \ .
\ee
By contracting these equations with $i_m i_k$ one gets \cite{ISm1}
\be
X \chi \w \df\Phi \heq{H} 0 \quad \textrm{and} \quad X \chi \w \df\Psi \heq{H} 0 \ .
\ee
From these two equations it follows that at any point of $H[\chi]$, $\df\Phi$ and $\df\Psi$ are either either zero or
proportional to $\chi_a$ except possibly at points where $X = 0$ (note that by definition we have $\df\Phi = 0$ and $\df\Psi = 0$ at the points where $\chi^a = 0$). Therefore, for any tangent vector $Z^a \in TH[\chi]$,
\be
i_Z \df\Phi \heq{H} 0 \quad \textrm{and} \quad i_Z \df\Psi \heq{H} 0 \ ,
\ee
so that $\Phi$ and $\Psi$ are constant over $H[\chi]$ except possibly at the points where rotation axis intersect the horizon $H[\chi]$ (where $X = 0$). Since the electromagnetic field 2-form $F_{ab}$ is by assumption nonsingular on the horizon $H[\chi]$ we know that potentials are continuous on $H[\chi]$ and thus the conclusion about the constancy can be extended to all points of the horizon.
}

\vspace{20pt}

\section{Uniqueness for weak fields}

The problem of uniqueness of black hole solutions to Einstein-Maxwell's equations is highly nontrivial, attacked with arsenal of techniques, yet resulting in surprisingly elegant no-hair theorems (for a comprehensive review see \cite{HCC}). On the other hand, a considerably simpler problem is the one of electromagnetic \emph{test} fields, those which are not sufficiently strong to affect the background metric and whose contribution to the energy-momentum tensor can be neglected. Uniqueness of such solutions to Maxwell's equations on the background of curved spacetime has been proven in the cases of static and stationary axisymmetric spacetimes \cite{Wald74,ME03}. Here we shall demonstrate that uniqueness can be established for a larger class of stationary spacetimes.

\bigskip

\btm
Let spacetime $(M,g_{ab})$ be a solution to Einstein's equation with nonnegative cosmological constant, $\Lambda \ge 0$, admitting a Killing vector field $\xi^a$ which is timelike in simply connected domain of outer communications $D \subset M$. Furthermore, let this spacetime contains electric and magnetic currents satisfing conditions (\ref{eq:ccm})--(\ref{eq:cce}) and electromagnetic test field $F_{ab}$, invariant under the action of the Killing vector field $\xi^a$ and nonsingular on $D$. Then the corresponding electric scalar potential $\Phi$ and the magnetic scalar potential $\Psi$, which satisfy Dirichlet boundary conditions on each component of the boundary $\dd D$, are uniquelly determined in $D$ if at each point of $D$ at least one of the conditions,
\be\label{eq:omegaphipsi}
\omega_a = 0 \quad \textrm{or} \quad \Psi = 0 \quad \textrm{or} \quad \Phi = 0
\ee
is satisfied.
\etm

\medskip

\noindent
Note that the validity of conditions (\ref{eq:omegaphipsi}) at all points of $D$ corresponds to, respectfully, static, purely electric and purely magnetic case. The choice of Dirichlet boundary conditions is natural in astrophysical setting, since the scalar potentials are constant on black hole horizons and asymptotically vanish at large distances when the currents have compact support.

\bigskip

\prf{
Let $\{\Phi^{(1)},\Psi^{(1)}\}$ and $\{\Phi^{(2)},\Psi^{(2)}\}$ be two solutions to Maxwell's equations (\ref{eq:Maxspot1})--(\ref{eq:Maxspot2}) with identical boundary conditions. Then the scalars $\phi \equiv \Phi^{(2)} - \Phi^{(1)}$ and $\psi \equiv \Psi^{(2)} - \Psi^{(1)}$ are solutions to vacuum (source-free) Maxwell's equations, 
\begin{eqnarray}
\Delta \phi - \frac{1}{N}\,\Big[ (\df N|\df\phi) + (\omega|\df\psi) \Big] & = 0 \label{eq:vMax1}\\
\Delta \psi - \frac{1}{N}\,\Big[ (\df N|\df\psi) - (\omega|\df\phi) \Big] & = 0 \label{eq:vMax2}
\end{eqnarray}
with vanishing boundary conditions. As an immediate consequence of the Stokes' theorem we have the following identity for all sufficiently differentiable scalars $f$ and $g$ defined on $D$,
\be\label{eq:green}
\int_D *\Big[ f \Delta g + (\df f|\df g) \Big] = \int_{\dd D} *(f\,\df g)
\ee
If we insert $f = g = \phi$ and $f = g = \psi$, use Maxwell's equation (\ref{eq:vMax1})--(\ref{eq:vMax2}) and boundary conditions, we shall obtain two equations,
\be\label{eq:midint1}
\int_D *\left[ (\df\phi|\df\phi) + \frac{\phi}{N}\,\Big( (\df N|\df\phi) + (\omega|\df\psi) \Big) \right] = 0
\ee
\be\label{eq:midint2}
\int_D *\left[ (\df\psi|\df\psi) + \frac{\psi}{N}\,\Big( (\df N|\df\psi) - (\omega|\df\phi) \Big) \right] = 0
\ee
Using basic identities for Killing vectors \cite{Heusler}, it can be shown that
\be\label{eq:deltadN}
\delta \left( \frac{\df N}{N} \right) = -\frac{(\omega|\omega) + 2R(\xi,\xi)}{N^2} \ .
\ee
The middle term under both integrals, (\ref{eq:midint1}) and (\ref{eq:midint2}), can be rewritten using partial integration and identity (\ref{eq:deltadN})
\be\label{eq:parcint}
\int_D *\,\frac{\gamma}{N}\,(\df N|\df\gamma) = \int_{\dd D} \frac{*\df N}{2N}\,\gamma^2 + \int_D *\,\frac{(\omega|\omega) - R(\xi,\xi)}{2N^2}\,\gamma^2
\ee
where $\gamma$ stands for $\phi$ and $\psi$. The first term on the rhs of (\ref{eq:parcint}) vanishes due to boundary conditions, but more careful treatment is required at horizons, where $N = 0$. Here one can use L'H\^opital rule to see that the ratio $\gamma/N$ remains bounded on nondegenerate horizons, while for most cases of degenerate (extremal) horizons, the same conclusion can be drawn by repeated use of L'H\^opital rule. Now, using Einstein's equation twice contracted with $\xi^a$, $R(\xi,\xi) = N\Lambda$, we have
\be
\int_D *\left[ (\df\phi|\df\phi) + \frac{(\omega|\omega) - N\Lambda}{2N^2}\,\phi^2 + \frac{\phi}{N}\,(\omega|\df\psi) \right] = 0
\ee
\be
\int_D *\left[ (\df\psi|\df\psi) + \frac{(\omega|\omega) - N\Lambda}{2N^2}\,\psi^2 - \frac{\psi}{N}\,(\omega|\df\phi) \right] = 0
\ee
The final assumption in the theorem allows us to dispose of the third terms under both integrals. From the assumptions that $\xi^a$ is timelike in $D$ and $\Lambda \ge 0$, it follows that $-N\Lambda \ge 0$. Furthermore, one immediate consequence of the definitions of the scalar potentials and the twist is that
\be
i_\xi \df\phi = i_\xi \df\psi = i_\xi \omega = 0 \ ,
\ee
so that the corresponding vector fields are spacelike and thus
\be
(\df\phi|\df\phi) \ge 0 \qqd (\df\psi|\df\psi) \ge 0 \qqd \textrm{and} \quad (\omega|\omega) \ge 0 \ .
\ee
In conclusion, we are left with the sum of positive semidefinite integrands, from where it follows that $\df\phi = \df\psi = 0$ on $D$, and thus $\phi$ and $\psi$ are constant on $D$. Since both $\phi$ and $\psi$ are zero on the boundary $\dd D$, the claim follows.
}

\bigskip

One has to be careful in application of the previous theorem when the spacetime contains the ergoregion $\mathcal{E}$. In such cases there will be a subset of the domain of outer communications where a timelike Killing vector field might become spacelike. Also, the ergosurface, that is the boundary of $\mathcal{E}$, is generally not an equipotential hypersurface. However, stationary axisymmetric and even helical spacetimes \cite{UGM} admit a Killing vector field appropriate for this theorem.

\bigskip

Let us return to the final assumptions from the theorem. In an attempt to relax them, one might try to observe the sum of the integrals,
$$\fl \int_D *\Big[ (\df\phi|\df\phi) + (\df\psi|\df\psi) + \frac{(\omega|\omega) - N\Lambda}{2N^2}\,(\phi^2 + \psi^2) \, +$$
\be
+ \, \frac{\phi}{N}\,(\omega|\df\psi) - \frac{\psi}{N}\,(\omega|\df\phi) \Big] = 0
\ee
Obviously, the claim would hold if the last two terms together would be a positive semidefinite function on $D$. However, it is difficult to see why would such an inequality hold in general. Another path towards the generalization of the uniqueness would be to recast Maxwell's equations into the form of a first order symmetric hyperbolic system of differential equations (see e.g.~\cite{Fried91} and Appendix A in \cite{Racz13}), so that the well-posedness is guaranteed.

\vspace{20pt}

\section{When can we utilise Einstein's equation as a Swiss army knife?}

One of the important consequences of the diffeomorphism invariance of the physical theory is the vanishing of the divergence of the energy-momentum tensor. This property can be broken by quantum effects due to eventual presence of diffeomorphism anomaly, and in such cases it can point to possible inconsistencies \cite{Bertlmann}, or can be used within some effective description of natural phenomena \cite{RW05}. However, at the classical level, these complications are always absent. It has been noticed that covariant conservation of energy-momentum is sometimes enough to derive complete equations of motion for matter fields, or at least some part of them. In order to better understand the background of such an assertion, let us carefully examine this relation from the perspective of the calculus of variations.

\bigskip

We assume that the total action $S = S_G + S_M$ consists of gravitational part $S_G[g_{ab}]$ and matter part $S_M[g_{ab},\psi]$, which are both functionally differentiable (see \cite{Wald}, Appendix E). Here we are using the abstract symbol $\psi$ with supressed indices to denote some general matter field. The energy-momentum tensor $T_{ab}$ is introduced via
\be
T_{ab} = -\frac{2}{\sqrt{-g}}\,\frac{\delta S_M}{\delta g^{ab}}
\ee
Let $\{g_{ab};\psi\}_\lambda$ be a smooth 1-parameter family of field configurations (with appropriate boundary conditions). Then the derivative of functional $S$ can be split as
\be\label{eq:varS}
\frac{\df S}{\df\lambda} = \int \frac{\delta S_G}{\delta g^{ab}}\,\delta g^{ab} + \int \frac{\delta S_M}{\delta g^{ab}}\,\delta g^{ab} + \int \frac{\delta S_M}{\delta\psi}\,\delta\psi
\ee
where the summation over all possible indices of the fields is assumed in the last term. If the fields $\{g_{ab};\psi\}_{\lambda_0}$ extremize total action $S$ for some value $\lambda_0$ of the parameter, so that $(\df S/\df\lambda)(\lambda_0) = 0$, we recover the gravitational and the matter equations of motion. Instead of looking at general variations we can specialize to 1-parameter family of diffeomorphisms $f_\lambda : M \to M$. We know that for such variations $\delta g_{ab} = \Lie_v g_{ab}$ for some vector field $v^a$. Using this one can prove \cite{Wald} that

\medskip

\begin{itemize}
\item[a)] As a consequence of invariance of the gravitational part of the action $S_G$ with respect to diffeomorphisms we have equality
\be
\int (\nabla^a E_{ab}) v^b \bm{\epsilon} = 0
\ee
which implies that $E_{ab}$, the functional derivative of the $S_G$ with respect to metric, is divergence free,
\be\label{eq:BianE}
\nabla^a E_{ab} = 0 \ .
\ee
More concretely, in the case of Einstein's tensor $E_{ab} = G_{ab}$, this equation is the twice contracted Bianchi identity,
\be\label{eq:BianG}
\nabla^a G_{ab} = 0 \quad ;
\ee

\medskip

\item[b)] As a consequence of invariance of matter part of the action $S_M$ with respect to diffeomorphisms \emph{and} assuming that $\psi$ satisfies the matter field equations, we have equality
\be
\int (\nabla^a T_{ab}) v^b \bm{\epsilon} = 0 \ ,
\ee
which implies the ``covariant conservation'' of the energy-momentum tensor $T_{ab}$,
\be\label{eq:BianT}
\nabla^a T_{ab} = 0 \ .
\ee
\end{itemize}

\bigskip

Now we want to see what happens if we reverse the logic. A tensor $E_{ab}$, which serves as a ``left hand side'' of the gravitational field equation $E_{ab} = 8\pi T_{ab}$, is usually chosen by construction to satisfy equation (\ref{eq:BianE}) as a \emph{mathematical identity}. This means that the gravitational field equation itself implies (\ref{eq:BianT}) and from here it follows that for all diffeomorphisms induced by the vector field $v^a$ we have
\be
\int \frac{\delta S_M}{\delta\psi}\,\delta_v\psi = 0 \ .
\ee
Using arbitrariness of the vector field $v^a$, this equation allows us to deduce \emph{only 4} components of the matter equations of motion, $\delta S_M/\delta \psi = 0$. These 4 components directly correspond to 4 components of the equation (\ref{eq:BianT}). So, in all cases where matter field $\psi$ has more than 4 components, deduction of equations of motion solely from the Bianchi identity is \emph{not} generally possible. In order to reduce the number of independent equations of motion for the matter field $\psi$, one can restric the analysis to the fields with some additional symmetries.

\bigskip

The most simple example is the one of real scalar field (see e.g.~\cite{MTW}, Exercise 20.9.) with Lagrangian density
\be
\mathscr{L} = -\sqrt{-g} \left( \frac{1}{2}\,g^{ab} \nab{a}\phi\,\nab{b}\phi + V(\phi) \right) \ ,
\ee
where $V(\phi)$ is a scalar potential which may include the mass term $V_{\mathrm{mass}}(\phi) = m^2\phi^2/2$. The corresponding equation of motion is the Klein-Gordon equation,
\be
\Box \phi - V'(\phi) = 0 \ ,
\ee
where $\Box = \nab{a}\nabla^a$ denotes the usual D'Alembertian operator. The energy-momentum tensor is given by
\be
T_{ab}^{(\phi)} = \nab{a}\phi\,\nab{b}\phi - g_{ab} \left( \frac{1}{2}\,g^{cd} \nab{c}\phi\,\nab{d}\phi + V(\phi) \right)
\ee
and its divergence
\be
\nabla^a T_{ab}^{(\phi)} = \Big( \Box\phi - V'(\phi) \Big) \nab{b}\phi \ .
\ee
Assuming that the scalar field $\phi$ is not constant, Klein-Gordon equation immediately follows from (\ref{eq:BianT}). In a case when $\nab{a}\phi$ vanishes on some subset of the spacetime, a more careful treatment is needed. For example, if the set
\be
S = \left\{ \, p \in M \, : \, \nab{a}\phi(p) = 0 \, \right\}
\ee 
is a hypersurface, then we can extend the values of $\phi$ to the points of $S$, under assumption that $\phi : M \to \rr$ is a continuous map.

\bigskip

More subtle examples are those involving nonminimally coupled scalar field. For example, if the Lagrangian density is given by \cite{Faraoni},
\be
\mathscr{L} = -\sqrt{-g} \left( \frac{1}{2}\,g^{ab} \nab{a}\phi \nab{b}\phi + V(\phi) - \frac{1}{2}\,F(\phi) R \right)
\ee
with some arbitrary function $F$ and Ricci scalar $R$, then the presence of $F(\phi) R$ term prevents clear-cut splitting of the action into ``gravitational'' and ``matter'' parts. Here, equations of motion are given by
$$\fl F(\phi) G_{ab} = \nab{a}\phi \nab{b}\phi - g_{ab} \left( \frac{1}{2}\,\nab{c}\phi \nabla^c\phi + V(\phi) \right) +$$
\be
+ \Big( \nab{b}\nab{a} - g_{ab}\,\Box \Big) F(\phi)
\ee
\be\label{eq:FRKG}
\Box\phi + \frac{1}{2}\,F'(\phi) R - V'(\phi) = 0
\ee
In order to cast the gravitational equation into the form reminiscent of Einstein equation, it is customary to (re)define energy-momentum tensor,
$$\fl G_{ab} = T_{ab}^{(\mathrm{eff})} \equiv \frac{1}{F(\phi)} \left[ \nab{a}\phi \nab{b}\phi - g_{ab} \left( \frac{1}{2}\,\nab{c}\phi \nabla^c\phi + V(\phi) \right) + \right.$$
\be\label{eq:GTphi}
+ \Big( \nab{b}\nab{a} - g_{ab}\,\Box \Big) F(\phi) \Bigg]
\ee
It has been noticed in the recent paper \cite{CarDun} that, assuming that the scalar field is not constant $\nab{a}\phi \ne 0$, generalized Klein-Gordon equation (\ref{eq:FRKG}) follows from the Bianchi identity,
$$\fl 0 = \nabla^b G_{ab} = \nabla^b T_{ab}^{(\mathrm{eff})} = \frac{1}{F(\phi)} \left( \Box\phi + \frac{1}{2}\,F'(\phi) R - V'(\phi) \right) \nab{b}\phi \, +$$
\be
+ \Big( G_{bc} - T_{bc}^{(\mathrm{eff})} \Big) F'(\phi) \nabla^c\phi \ ,
\ee
and additional use of the gravitational equation (\ref{eq:GTphi}).

\bigskip

Another simple example is ideal fluid, described by two scalars, energy density $\rho$ and pressure $p$, and 4-velocity $u^a$ of the ``fluid element''. These 5 degrees of freedom are usually reduced to 4 by additional assumption of functional relation between energy density and pressure, $p = f(\rho)$, known as the equation of the state. Naive counting of the number of degrees of freedom correctly suggests that the equations of motion, relativistic Euler equations for ideal fluid, can be deduced from the Bianchi identity (see e.g.~\cite{Wald}, page 69).

\bigskip

Unfortunately, the list of successful examples is quickly exausted. Nonefficiency of the procedure has already been encountered in the case of complex scalar field, as well as for the massless Weyl spinor field and the massive Dirac field. The obstacle in these spin-half cases comes from the fact that the covariant conservation of the energy-momentum tensor corresponds to a second order partial differential equation, whereas the Weyl and Dirac equations are first order partial differential equations (see \cite{PR}, chapter 5.8). The case of coupled Einstein-Dirac system of equations has been analysed by Friedrich and Rendall in \cite{FR00}, where they have treated it as a Cauchy problem.

\vspace{20pt}

\section{From Einstein and Bianchi to Maxwell}

The electromagnetic field 2-form $F_{ab}$ in 4 dimensional spacetimes has 6 components. This already suggests that we have to make some additional assumptions in order to recover Maxwell's equations from the Bianchi identity. One pragmatic solution has been proposed in \cite{MTW}, where the authors simply assume one half of the Maxwell's equations and then derive the other half from the Bianchi identity. On the other hand, R\'acz has shown \cite{Racz93} that presence of one Killing vector field is enough to recover \emph{complete} vacuum Maxwell's equations, except in some degenerate cases. We shall now demonstrate how this argument can be expanded to a nonvacuum case, and how the current conditions play a crucial role in this procedure.

\bigskip

Let $(M,g_{ab},F_{ab})$ be a spacetime admitting a Killing vector field $\xi^a$, which is non-null (i.e. $N \ne 0$) on some open set $O$, and containing electromagnetic field $F_{ab}$ which is invariant under the action of the vector field $\xi^a$ and nonsingular in $O$. Note that from this assumption follows that $\Phi$ and $\Psi$ are at least continuous, and thus their values can be extrapolated to points where $N = 0$. Electromagnetic energy-momentum tensor is given by
\be
T_{ab}^{(\mathrm{em})} = \frac{1}{4\pi} \left( \tensor{F}{_a^c} F_{bc} - \frac{1}{4}\,g_{ab}\,F_{cd} F^{cd} \right)
\ee
and its divergence can be evaluated with help of Maxwell's equations (\ref{eq:MaxF}),
\be
\nabla^a T_{ab}^{(\mathrm{em})} = \je^a F_{ab} - \jm^a\,{*F}_{ab}
\ee
Assuming that current conditions are satisfied, using equations (\ref{eq:jxijxi}), rhs can be written as
\be\label{eq:BianEM}
\nabla^a T_{ab}^{(\mathrm{em})} = -\frac{(\xi|\je)}{N}\,E_b - \frac{(\xi|\jm)}{N}\,B_b
\ee
Conversely, the Bianchi identity (\ref{eq:BianG}) applied to Einstein's equation (\ref{eq:EE}) (or more generally, identity (\ref{eq:BianE}) applied to the general gravitational field equation $E_{ab} = 8\pi T_{ab}$) with the \emph{total} energy-momentum tensor imply (\ref{eq:BianEM}) via current conditions. This form of the divergence of the electromagnetic energy-momentum tensor will turn out to be particularly convenient for the deduction of Maxwell's equations. On the other hand, the electromagnetic energy-momentum tensor can be decomposed with respect to electric and magnetic fields (see \cite{Heusler,Racz93}) as
$$4\pi T_{ab}^{(\mathrm{em})} = N^{-2} \left[ \Big( E_c E^c + B_c B^c \Big) \left( \xi_a \xi_b - \frac{N}{2}\,g_{ab} \right) + 2\xi_{(a} S_{b)} \right] +$$
\be
+ N^{-1} (E_a E_b + B_a B_b)
\ee
where we have introduced Poyting 1-form $S_a$,
\be
S \equiv *(\xi \w E \w B)
\ee
Assuming that the current conditions are satisfied, we can introduce scalar potentials and express the divergence of the electromagnetic energy-momentum tensor as
$$\fl 4\pi \nabla^a T_{ab}^{(\mathrm{em})} = N^{-1} \left( \Delta\Phi - \frac{1}{N}\,(\df\Phi|\df N) \right) \nab{b}\Phi \, +$$
\be\label{eq:divT1}
+ N^{-1} \left( \Delta\Psi - \frac{1}{N}\,(\df\Psi|\df N) \right) \nab{b}\Psi + \nabla^a \left( 2 N^{-2} \xi_{(a} S_{b)} \right)
\ee
Further analysis can be divided to the following cases

\vspace{20pt}

a) If $E \w B = 0$ then either $E_a = 0$ or $B_a = 0$ or $E_a \sim B_a$ (in all cases $S_a = 0$). As we shall see, these are the degenerate cases, where the deduction of Maxwell's equations is not efficient. Let us first assume that $E_a = 0$ and $B_a \ne 0$. This means that $\Phi$ is constant,
\be
4\pi \nabla^a T_{ab}^{(\mathrm{em})} = N^{-1} \left( \Delta\Psi - \frac{1}{N}\,(\df\Psi|\df N) \right) \nab{b}\Psi
\ee
Comparison with (\ref{eq:BianEM}) recovers the Maxwell's equation for the magnetic scalar potential $\Psi$. However, we are missing the remaining Maxwell's equation,
\be
(\omega|\df\Psi) = 4\pi N (\xi|\je)
\ee
which cannot be recovered in this way. The case when $B_a = 0$ and $E_a \ne 0$ is completely analogous. Finally, if $E_a \sim B_a$ holds, then
$$\fl 4\pi \nabla^a T_{ab}^{(\mathrm{em})} = N^{-1} \left( \Delta\Phi - \frac{1}{N}\,(\df\Phi|\df N) \right) \nab{b}\Phi \, +$$
\be
+ N^{-1} \left( \Delta\Psi - \frac{1}{N}\,(\df\Psi|\df N) \right) \nab{b}\Psi
\ee
Although we recognise Maxwell's equations inside of this expression, we cannot separate them since $\df\Phi$ and $\df\Psi$ are, by assumption, collinear. At best one can say that the Bianchi identity, together with one half of the Maxwell's equations implies the other half.

\bigskip

b) If $E \w B \ne 0$ and $\xi^a$ is a timelike vector then the vectors in the set $\{\xi^a,E^a,B^a,S^a\}$ are linearly independent, and thus form a well defined vector basis (namely, in this case all three vectors, $E^a$, $B^a$ and $S^a$, have to be spacelike). If $\xi^a$ is spacelike, then it may happen that $S^a$ is null,
\be
(S|S) = N \Big[ (E|B)^2 - (E|E)(B|B) \Big] = 0 \ ,
\ee
from where it follows that
\be
S \w E \w B = 0 \ .
\ee
This means that $S^a$ is parallel to either $E^a$ or $B^a$. Let us denote the parallel one by $C^a$ and the other one by $\widehat{C}^a$. In this case one can choose different, pseudo-orthogonal basis $\{\xi^a,E^a,B^a,\sigma^a\}$, where $\sigma^a$ is the unique null vector field, such that $(\sigma|C) = -1$ and $(\sigma|\xi) = (\sigma|\widehat{C}) = 0$. Using appropriate base to decompose the last term on the rhs of (\ref{eq:divT1}), it can be shown \cite{Racz93} that the divergence of the energy-momentum tensor can be written in the form
$$\fl 4\pi \nabla^a T_{ab}^{(\mathrm{em})} = N^{-1} \left\{ \Delta \Phi - \frac{1}{N}\,\Big[ (\df N|\df\Phi) + (\omega|\df\Psi) \Big] \right\} \nab{b}\Phi \, +$$
\be\label{eq:finaldivT}
+ N^{-1} \left\{ \Delta \Psi - \frac{1}{N}\,\Big[ (\df N|\df\Psi) - (\omega|\df\Phi) \Big] \right\} \nab{b}\Psi + N^{-1} \nabla^a\!\left( N^{-1} S_a \right) \xi_b
\ee
Finally, linear independence of the 1-forms $\{\nab{a}\Phi,\nab{a}\Psi,\xi_a\}$ allows us to compare (\ref{eq:finaldivT}) with (\ref{eq:BianEM}) and deduce the Maxwell's equations (\ref{eq:Maxspot1})--(\ref{eq:Maxspot2}). In addition, we also recover the conservation of the (Noether) current $J^a = N^{-1} S^a$,
\be
\nab{a} J^a = 4\pi\,\xi^b \nabla^a T_{ab}^{(\mathrm{em})} = 0
\ee

\vspace{20pt}

\section{Final remarks}

The very existence of the electromagnetic scalar potentials, along with their basic properties in the spacetimes with symmetries, is tightly connected with the current conditions (\ref{eq:ccm})--(\ref{eq:cce}). This provides us with a new insight about the assumptions used in the generalizations of the ``zeroth law'' of black hole electrodynamics: The same staticity and circularity conditions used in proofs are simultaneously a guarantee that the scalar potentials are locally well defined.

\bigskip

It would be interesting to investigate to what extent could the scalar potentials be of use for the force-free electrodynamics \cite{BGJ13,GJ14}. In this setting one deals with the electromagnetic test field, subject to the \emph{force-free condition},
\be\label{eq:ffe}
\je^a F_{ab} = 0 \ ,
\ee 
and absent of the magnetic monopoles, $\jm^a = 0$. If, in addition, the electric current $\je^a$ satisfies condition (\ref{eq:cce}), then (\ref{eq:ffe}) corresponds to the choice of purely magnetic field and Maxwell's equations are reduced to one ordinary differential equation for the magnetic scalar potential $\Psi$.

\bigskip

A brief analysis of the deduction of the equations of motion for matter from the twice contracted Bianchi identity has revealed the limitations of such a procedure. Nevertheless, the Bianchi identity can provide useful information, as can be seen e.g.~in the general $1+n$ decomposition of Einstein's equation \cite{Racz14}. Furthermore, electromagnetic field in the spacetimes with symmetries has a reduced number of degrees of freedom, and we have demonstrated that this allows one to derive complete nonvacuum Maxwell's equations from the Bianchi identity, except in some degenerate cases.

\bigskip

All the results presented here rely on the presence of the \emph{exact} symmetries of the spacetime and the fields. In the absence of any Killing vector field it will be difficult in practice, although not impossible in principle, to find the appropriate vector field $X^a$, such that $E_a(X)$ and $B_a(X)$ are closed 1-forms.

\vspace{30pt}

\ack

The author would like to thank the hospitality of Wigner Research Center of Physics, and especially to Istv\'an R\'acz for many enlightening discussions and encouragement during the preparation of this paper. This work was partially supported by the Croatian Ministry of Science, Education and Sport under the contract no.~119-0982930-1016 and partially by the grant ``Natje\v caj za mobilnost istra\v ziva\v ca''.

\vspace{30pt}

\appendix

\section{Basic formulae of differential geometry}

Let $(M,g_{ab})$ be a $m$-dimensional Lorentzian manifold and $\Omega^p(M)$ space of $p$-forms defined on it. For every $\omega \in \Omega^p(M)$ we define Hodge operator $* : \Omega^p \to \Omega^{m-p}$, contraction $i_X : \Omega^p \to \Omega^{p-1}$ with a vector field $X^a$, and exterior derivative $\df : \Omega^p \to \Omega^{p+1}$, written in abstract index notation as
\be
(*\omega)_{a_{p+1} \cdots a_m} = \frac{1}{p!}\,\omega_{a_1 \cdots a_p} \tensor{\epsilon}{^{a_1}^{\cdots}^{a_p}_{a_{p+1}}_{\cdots}_{a_m}}
\ee
\be
(i_X \omega)_{a_1 \cdots a_{p-1}} = X^b \omega_{b a_1 \cdots a_{p-1}}
\ee
\be
(\df\omega)_{a_1 \cdots a_{p+1}} = (p+1) \nab{[a_1} \omega_{a_2 \dots a_{p+1}]}
\ee
For any $\alpha\in\Omega^p(M)$ and $\beta\in\Omega^q(M)$ we define \emph{wedge product} $\w : \Omega^p(M) \times \Omega^q(M) \to \Omega^{p+q}(M)$ as
\be
(\alpha \w \beta)_{a_1 \cdots a_p b_1 \cdots b_q} = \frac{(p+q)!}{p!q!}\,\alpha_{[a_1 \cdots a_p} \beta_{b_1 \cdots b_q]}
\ee
The two following identities hold for every $\omega \in \Omega^p(M)$,  
\be
{**}\,\omega = (-1)^{p(m-p)+1}\,\omega
\ee
\be\label{eq:iXsalpha}
i_X *\omega = *(\omega \w X)
\ee
Hodge dual of a $0$-form $f$ is directly related to the volume form $\bm{\epsilon} \in \Omega^m(M)$,
\be
*f = f\,\bm{\epsilon} \ .
\ee
For any pair of $p$-forms $\alpha$ and $\beta$, their \emph{inner product} is defined by
\be
(\alpha|\beta) = \frac{1}{p!}\,\alpha_{a_1 \cdots a_p} \beta^{b_1 \cdots b_p} \ ,
\ee
which satisfies
\be
(\alpha | \beta) = -\,*(\alpha \w *\beta)
\ee
and
\be
(i_X \alpha | \beta) = (\alpha | X \w \beta) \ .
\ee
Additional derivative operators are coderivative $\delta : \Omega^p(M) \to \Omega^{p-1}(M)$ and Laplace-Beltrami operator $\Delta : \Omega^p(M) \to \Omega^p(M)$, defined as
\be
\delta = (-1)^{m(p+1)+1}\,{*\,d\,*} \qqd \Delta = d \delta + \delta d \ .
\ee
Note that for every $0$-form $f$ we have $\delta f = 0$ and $\Delta f = \Box f = \nab{a} \nabla^{a} f$. Lie derivative of $\omega \in \Omega^p(M)$ with respect to a vector field $X^a$ satisfies Cartan's identity,
\be\label{eq:Cartan}
\Lie_X \omega = i_X \df\omega + \df i_X \omega \ .
\ee
Important fact is that the Lie derivative with respect to a Killing vector field $\xi^a$ commutes with Hodge dual,
\be
\Lie_\xi\,{*\,\omega} = *\,\Lie_\xi\,\omega
\ee

\vspace{20pt}

\section*{References}

\nocite{*}

\bibliographystyle{unsrt}
\bibliography{emspot}

\end{document}